\documentclass[aps,twocolumn,pra,floatfix,groupedaddress,nofootinbib,notitlepage,showpacs,superscriptaddress]{revtex4-1}
\usepackage{amsmath}
\usepackage{times,bm,bbm,bbold,graphicx,graphics,amssymb,amsmath,amsfonts,dsfont,hyperref,color}
\usepackage{mathrsfs,cancel}
\usepackage[utf8]{inputenc}
\usepackage{graphicx}
\usepackage{soul,xcolor}
\setstcolor{red}
\definecolor{mycolor}{rgb}{0.1, 0.1, 0.7}
\usepackage{dsfont}
\usepackage{url}
\usepackage{soul}
\usepackage[vlined,ruled]{algorithm2e}

\DeclareFontFamily{OT1}{pzc}{}
\DeclareFontShape{OT1}{pzc}{m}{it}%
{<-> s * [1.25] pzcmi7t}{}
\DeclareMathAlphabet{\mathpzc}{OT1}{pzc}%
{m}{it}
\hypersetup{
	plainpages=true,
	breaklinks=true,
	hypertexnames=false,
	pageanchor=true,
	colorlinks=true,
	linkcolor={mycolor},
	citecolor={mycolor},
	urlcolor={mycolor},
	anchorcolor={black}
}

\begin{document}
	\title{How well can we guess the outcome of  measurements of  non-commuting observables?}
	\author{Maryam Khanahmadi}
	\email{m.khanahmadi@phys.au.dk}
	\affiliation{Center for Complex Quantum systems, Department of Physics and Astronomy, University of Aarhus, Ny Munkegade 120, DK 8000 Aarhus C, Denmark } 
	\affiliation{Department of Physics, Institute for Advanced Studies in Basic Sciences, Zanjan 45137, Iran}
	
	\author{Klaus M{\o}lmer}
	\email{moelmer@phys.au.dk}
	\affiliation{Center for Complex Quantum systems, Department of Physics and Astronomy, University of Aarhus, Ny Munkegade 120, DK 8000 Aarhus C, Denmark } 
	\begin{abstract}
		According to Heisenberg's uncertainty relation, there is an ultimate limit to how precisely we may predict the outcome of position and momentum measurements on a quantum system. We show that this limit may be violated by an arbitrarily large factor if one aims, instead, to guess the unknown value of a past measurement. For experiments on a single quantum system, the precise assignment of past position and momentum measurement outcomes is accompanied by large uncertainty about their linear combinations, while we show that entanglement with an ancillary system permits accurate retrodiction of any such linear combination. Finally, we show that the  outcomes of experiments that jointly measure multiple linear combinations of position and momentum observables by means of ancillary probe particles can also be guessed with no formal lower limit. We present quantitative results for projective measurements and for generalized measurements where all components are prepared in Gaussian states. 
	\end{abstract}
	\pacs{
		03.65.Ta, 06.20.Dk
	}
	\date{\today}
	\maketitle
	
	\section{Introduction}
	The Heisenberg uncertainty relation (HUR) \cite{HUR} is regarded as a foundational insight and defining property of quantum physics. It is, colloquially, related to statements about the impossibility to measure a particle's position without affecting its momentum, or about the impossibility to simultaneously measure position $x$ and momentum $p$ of a particle. But, in fact the mathematical textbook relation, known as Heisenbergs uncertainty relation, neither concerns sequential nor simultaneous measurements of position and momentum, but only statements that can be made about measurements of either one of them. It follows directly from Born's rule for the probability distribution and the resulting statistical variance of the outcomes of measurements of any of the two observables. Calculating these quantities, $(\Delta x)^2$ and $(\Delta p)^2$, separately for any given quantum state, it was shown by Kennard in 1927 \cite{Kennard} that their product obeys what we now refer to as the Heisenberg uncertainty relation, 
	\begin{equation}\label{hur}
	\Delta x \Delta p \geq \hbar/2.
	\end{equation}
	To test the HUR, one should infer the variances by  making large numbers of position measurements and large numbers of momentum measurements on systems that are all identically prepared in the same state.
	
	Heisenberg famously discussed a similar, more qualitative, relationship governing the spatial resolution of optical microscopy and the magnitude of momentum kicks imparted on the particle under observation. The issue of  a physical disturbance of a quantum state by measurements and hence a purported difficulty to guess the outcome of subsequent measurements has been addressed in great detail \cite{Ozawa,busch}. Practical quantum optical schemes that are sensitive to both phase and amplitude changes of an electromagnetic field have been analyzed and found to obey similar inequalities as \eqref{hur} with extra terms and larger threshold values \cite{caves,arthursgoodman}. Schr{\"o}dinger \cite{Schrodinger} derived an uncertainty relation concerning the position observable at different times, and also for that case, the question of determining the value measured at one or another time is very different from sequential probing at both instances of time with the purpose, e.g., to determine changes due to the passage of a gravitational wave, \cite{woolley,Zander,caves,caves2,moller,tse}. The practical and fundamental limits to quantum sensing by multiple measurements are a matter of current research and crucial for the prospects of quantum enabled sensing technologies \cite{qsens}.
	
	In this article, we present analyses of the quantum mechanical uncertainty in different measurement scenarios. For all situations, we use conventional quantum mechanical formalism to determine the variance of the distribution of measurement outcomes around the values guessed from the knowledge about the system. In Sec. \ref{secII} we review elements of the quantum theory of measurements, needed for our analysis. In particular we provide the probabilities for the unknown outcomes of past measurements, conditioned on both prior and posterior observations of a quantum system \cite{aharonov,ABL,PQS}. In Sec. \ref{S:ss}, we show that when past measurements are concerned,  the outcome of position and momentum measurements can both be guessed with high precision in apparent violation of the HUR. In Sec. \ref{S:PAST}, we show that for suitable pre- and postselected entangled states of pairs of particles, it is possible to  ascertain the outcome of the measurement of any linear combination of the (dimension less) position and momentum observables of either one of the particles. In Sec. \ref{S:sn}, we demonstrate the theoretical possibility to perform -  and guess the outcome of - joint measurements of many such observables: a continuous variable version of the so-called Mean King's problem  \cite{vaidman,Mermin}. We supplement a previous analysis of that problem with idealized, perfect resolution and infinitely squeezed  states \cite{Aharonov,Botero} with more realistic Gaussian states. Sec. \ref{S:S}  provides a conclusion and an outlook.

	
	\section{Measurements in quantum mechanics}\label{secII}
	
	A quantum system  can at a given time $t$ be described by a wave function $\psi(t)$  or, more generally, by a density matrix $\varrho(t)$. Measurements of an observable, are ideally able to resolve all its eigenvalues which occur as  measurement outcome with probabilities given by the state overlap with the corresponding eigenstates. In somewhat more realistic descriptions, the system is coupled to another quantum system, the meter, and the true measurement is carried out on a meter degree of freedom, e.g., its position, which may have been displaced due to the interaction with the system observable. If the typical meter displacements exceed its position uncertainty, the measurement is equivalent to the measurement of the distinct eigenvalues of the system observables, but it may also only partially resolve these values.
	
	Formally, measurements can be described by a positive operator valued measure (POVM) which assigns operators $\lbrace \hat{M}_m\rbrace$ to each meter outcome $m$, satisfying the completeness relation $\sum _m \hat{M}_m^{\dagger} \hat{M}_m = \textbf{1}$. The probability of the measurement outcome  $m$ is
	\begin{align}\label{b}
	P(m,t) = \mathrm{Tr}(\hat{M}_m\varrho (t)\hat{M}_m^{\dagger}),   
	\end{align}
	and conditioned on that outcome, the system occupies the state $ \hat{M}_m \varrho \hat{M}_m^{\dagger}$, which may be subsequently normalized. If $\hat{M}_m$ are projection operators on the eigenstates of an  observable with eigenvalues distinguished by $m$, this yields the conventional Born rule and projection postulate. The formalism is readily generalized \cite{instruments} to account for cases where, e.g., the meter system is initially prepared in a mixed state or the measurement process is imperfect.   
	
	The POVM formalism yields a simple means to address the joint probability for the outcome of of consecutive measurements on a single quantum system. Let us address the possibility that the first measurement yields the value $m_1$ and the second measurement of the same or some other operator yields $m_2$ (we may readily incorporate unitary and dissipative evolution between the measurements, but for simplicity of notation we shall ignore such elements in this article). It follows from the above that 
	the joint probability $P(m_1,m_2)$ factors as the probability of first getting $m_1$ and then getting $m_2$, conditioned on the system occupying the state $\propto \hat{M}_{m_1} \varrho \hat{M}_{m_1}^{\dagger}$. Accounting properly for the normalization factors, this yields the joint probability,
	\begin{align}\label{bb}
	P(m_1,m_2) = \mathrm{Tr}(\hat{M}_{m_2}\hat{M}_{m_1}\varrho \hat{M}_{m_1}^{\dagger}\hat{M}_{m_2}^{\dagger}). 
	\end{align}
	This is an important relation, and it shows how the first measurement may for example be a sharp position measurement with outcome  $x \sim m_1$ which causes a back action that significantly alters the quantum state towards a position eigenstate with a large momentum uncertainty, while an unsharp position measurement would have yielded more fluctuating values for $x \sim m_1$ but led to a correspondingly minor disturbance of the state and hence a finite variance of the outcome of a subsequent momentum measurement of $p \sim m_2 $. The product of the variances of such a pair of measurements exceeds $\hbar$, i.e., it surpasses the usual HUR by at least a factor of two \cite{Kelly}. In quantum optics, a beam splitter permits separate  measurements of both quadratures of the quantized field and is known to display a similar doubling of the variance, known also as the half-quantum or 3 dB vacuum noise contribution \cite{caves3,arthursgoodman}.         
	
	While we may write the joint probability distribution as a product $P(m_1,m_2) = P(m_1)P(m_2|m_1)$ which  reflects the conditional evolution of the state by the first measurement, we may also formally define a conditional probability $P(m_1|m_2)$ through the relation, $P(m_1,m_2) = P(m_2)P(m_1|m_2)$. The quantity $P(m_1|m_2)$ is the probability that the first measurement yielded outcome $m_1$ conditioned on our knowledge that the later measurement yielded $m_2$.
	
	We can write this probability in the following way
	\begin{align}\label{pqs}
	P_P(m_1,t) = \frac{\mathrm{Tr}(\hat{M}_{m_1}\varrho (t)\hat{M}_{m_1}^{\dagger}E)}{\sum _{m_1'}\mathrm{Tr}(\hat{M}_{m_1'}\varrho (t)\hat{M}_{m_1'}^{\dagger}E)}
	\end{align}
	where $E = \hat{M}_{m_2}^{\dagger}\hat{M}_{m_2}$ is a positive operator (like $\varrho$) and the sum over $ m_1' $ includes all possible outcomes of the first measurement and normalizes the expression. Through \eqref{pqs} the pair of matrices $(\varrho, E)$ together permit calculation of the probabilities of the outcomes of any measurement conditioned on our knowledge about the preparation of the system and any later measurements on the system. Eq.\eqref{pqs} thus generalizes the conventional application of the quantum state $\varrho$ in Eq.\eqref{b} and qualifies the term Past Quantum State for the pair of matrices $(\varrho, E)$ \cite{PQS}. In \cite{PQS} we show that for systems subject to continuous monitoring, $\varrho$ and $E$ are both time dependent quantities, subject to stochastic master equations, while in this work, we shall merely assume the preparation of a given state $\varrho$ and the existence of a measurement protocol, abrupt or extended over time, that ensures the value of $E$ at the time $t$ right after the measurement for which we want to guess the outcome ($m_1$).

	Special cases of \eqref{pqs} include the ABL rule \cite{ABL} of projective measurements and the so-called weak values \cite{aharonov} which are obtained from the weighted average of the outcomes $m_1$ by weakly coupled meter systems. We also note related work by Hall \cite{Hall}, which shows how to maximize the knowledge about an observable $\hat{A}$, acquired in experiments by inclusion of prior information. In our notation, \cite{Hall} minimizes the statistical deviation between $\hat{A}$ and an optimally chosen function of the measurement effect operators, $\sum_{m_2} f(m_2) \hat{M}_{m_2}^{\dagger}\hat{M}_{m_2}$, and shows how this depends on the prior knowledge of the initial state in the form of $\varrho$.

	\section{Gaussian states and Gaussian measurements}\label{S:ss}
	The minimum uncertainty states of position and momentum observables are Gaussian wave functions in both variables, and we restrict the analysis of this work to position and momentum measurements for Gaussian states, i.e., for density matrices $\varrho$ corresponding to Gaussian Wigner functions $W_{\varrho}(x,p)$, with Gaussian marginal probability distributions, e.g., $P(x)=\int W_{\varrho}(x,p) dp$. These states are fully characterized by the mean values and second moments, of the position and momentum observables.    
	For a Gaussian quantum state $\varrho$ of a system with a single position and momentum coordinate, the covariance matrix is defined as
	\begin{align}\label{s4}
	\bm{\sigma} = \frac{1}{2}\mathrm{Tr}\big[ \lbrace (\hat{r}-\bar{r}),(\hat{r}-\bar{r})^{\mathrm{T}}
	\rbrace \varrho \big] 
	\end{align}
	where $\bm{\hat{r}}^{\mathrm{T}} = (\hat{x},\hat{p})$ represents the canonical operators with mean values $\bm{\bar{r}}^{\mathrm{T}} = ( \bar{x},\bar{p})$. We define the covariance matrix elements by
	$ \bm{\sigma} = \begin{bmatrix}
	\sigma _{x} &\sigma_{xp}\\\sigma_{xp}&\sigma_{p}
	\end{bmatrix}$.  Henceforth, we shall set $\hbar =1$, and assume rescaled dimensionless position and momentum variables. The HUR \eqref{hur} for $\Delta x = \sqrt{\sigma_x}, \Delta p = \sqrt{\sigma_p}$ implies that the covariance matrix elements obey $\sigma_x\sigma_p\geq 1/4$. 
	
	All operators have a Wigner function representation and we assume in the following that the effect matrix $E$ is formed by the POVM elements of a Gaussian measurement operation such as homodyne or heterodyne detection of field quadrature observables that have coupled linearly with the system observables $\hat{x}$ and $\hat{p}$. The operator $E$ then has a Gaussian Wigner distribution. While $E$ is a not a state in the same sense as $\varrho$, its Wigner function may thus also be fully characterized by its first and second moments, $\bm{\bar{r}}_E$ and $\bm{\sigma}_E$, respectively. .    
	
	\section{Past measurement outcomes can violate the Heisenberg Uncertainty Relation}\label{S:PAST}
	
	We shall now address the probability distribution of the outcome of a measurement of the position or any linear combination of the dimensionless position and momentum, $\hat{x}_\phi = \hat{x}\cos \phi + \hat{p}\sin \phi $, where we note the commutator relation $ [x_{\phi},x_{\phi+\pi/2}] = i $.  The observable $\hat{x}_\phi$ has eigenstates denoted by $|x_\phi\rangle$, and the outcome probability distribution is given by the marginal of the Wigner function, 
	$P(x_\phi)=  \langle x_\phi| \varrho |x_\phi\rangle = \int\mathrm{d}x_{\phi+\frac{\pi}{2}} W_{\varrho}(x_\phi,x_{\phi+\pi/2})$. The conventional HUR \eqref{hur} applies for the pair of observables  $x_{\phi},x_{\phi+\pi/2}$.
	
	\subsection{The uncertainty of retrodiction}
	
	We now turn to the variance of the outcomes of measurements of $\hat{x}_\phi$ conditioned on both prior and posterior information in the form of the operators $\varrho,E$. We employ Eq. \eqref{pqs}, and to this end we use the fact that the measurement of $\hat{x}_\phi$ is described by the POVM (projection) operators $\hat{M}_\phi = \hat{M}^{\dagger}_\phi = |x_\phi\rangle\langle x_\phi|$.
	Hence the outcome probability distribution for the projective position and momentum  measurements for a prepared Gaussian state and with a known posterior Gaussian measurement is given by
	\begin{align}\label{e1}
	\mathrm{P}_{P}(x_\phi,t) = \frac{\mathrm{Tr}(\hat{M}_\phi \varrho(t)\hat{M}^{\dagger}_\phi E(t))}{\int_{x_{\phi'}}\mathrm{Tr}(\hat{M}_{\phi'} \varrho(t)\hat{M}^{\dagger}_{\phi'} E(t))} \propto \langle x_\phi| \varrho|x_\phi\rangle\langle x_\phi| E|x_\phi\rangle 
	\end{align}
	where we used that $\hat{M}_\phi^2= \hat{M}_\phi,\ (\hat{M}^{\dagger}_\phi)^2 =
	\hat{M}^{\dagger}_\phi$.
	
	For the calculations we use that $\langle x_\phi| E|x_\phi\rangle$ is determined by the marginal of the  Wigner function the same way as $\langle x_\phi| \varrho|x_\phi\rangle$. For Gaussian states, the marginal distributions are Gaussian functions and so is their product. 
	
\begin{figure}[ht!]
		\includegraphics[scale=.45]{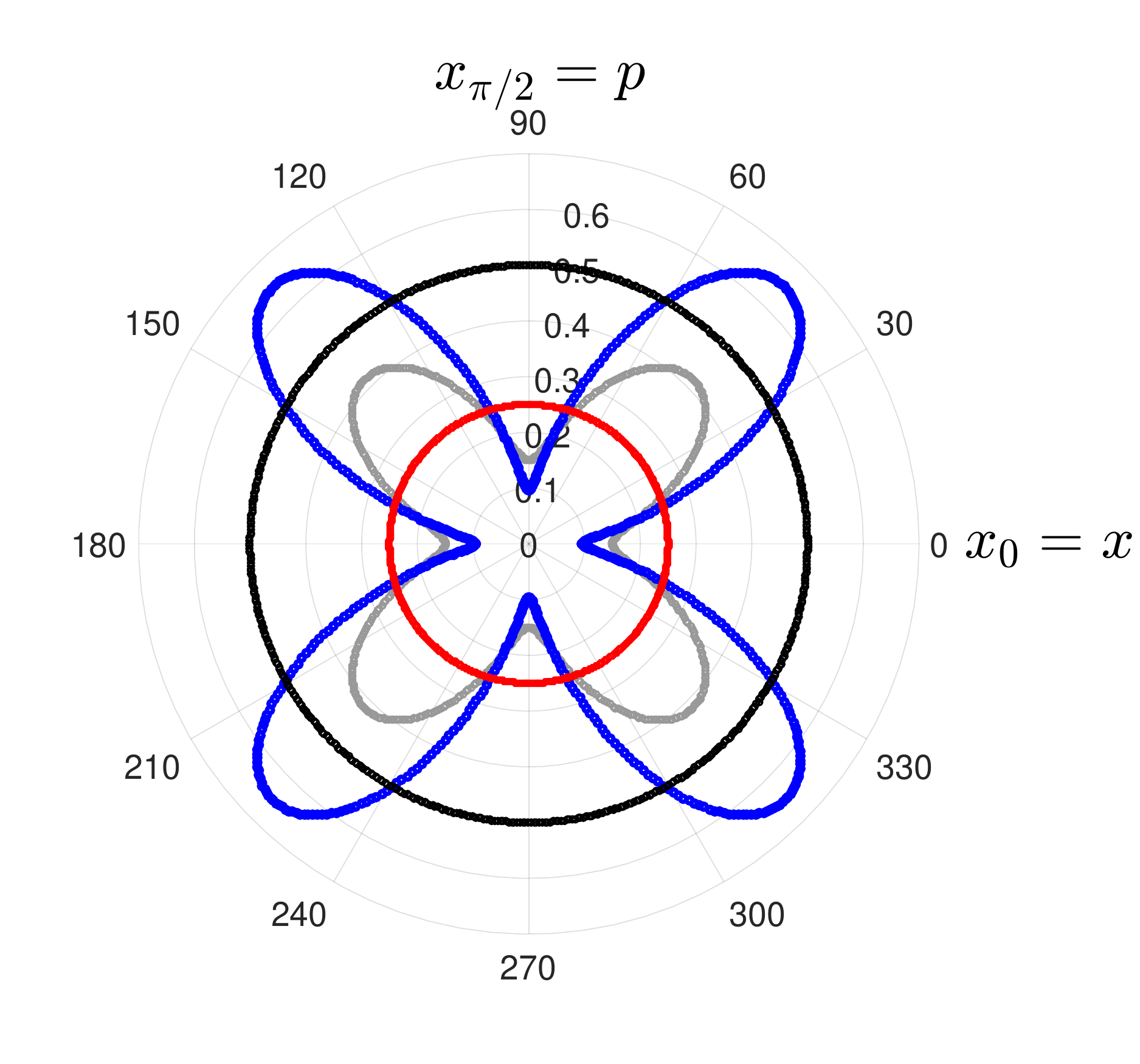}
		\caption{ Polar plot of the variance of the measurement outcomes of $x_\phi$ around their predicted or retrodicted most likely value. The outer black circle represents the conventional variance for a Gaussian minimum uncertainty state with equal variance in $x$ and $p$. The blue and gray curves show Eq. \eqref{ee3}, for a single system with $\varrho$ squeezed along $p$ and $E$ along $x$. We consider two diagonal covariance matrices for both $\varrho$ and $E$, where the blue and gray curves have the covariance matrix elements $\lbrace\sigma_{\varrho,x}= \sigma_{E,p}=5/2 ,\,\sigma_{\varrho,p}=\sigma_{E,x}=1/10\rbrace$ and $\lbrace\sigma_{\varrho,x}= \sigma_{E,p}=3/2 ,\,\sigma_{\varrho,p}=\sigma_{E,x}=1/6\rbrace$, respectively. The red circle is for the case where both $\varrho$ and $E$ correspond to  coherent states with $\lbrace\sigma_{\varrho,x}= \sigma_{E,p}= \sigma_{\varrho,p}=\sigma_{E,x}=1/2\rbrace$.}
		\label{f1}
	\end{figure}

	For the initial preparation of a coherent state which  fulfils the HUR with equality for all pairs of observables $x_{\phi},x_{\phi+\pi/2}$, and the final projection on a coherent state, the  product of the two distributions in Eq. \eqref{e1} yields a measurement outcome distribution with half the variances of the coherent state. This is illustrated with the inner red circle in Fig \ref{f1}. 
	
	In Ref. \cite{Jinglei}, it is shown that for arbitrary Gaussian $\varrho$ and $E$ for a single system, the variance of the  past probability $P_P(x_\phi)$ \eqref{e1} is given by
	\begin{align}\label{ee3}
	\sigma _P(x_\phi) = ( \frac{1}{\sigma_{\varrho,x_\phi}}+
	\frac{1}{\sigma_{E,x_\phi}})^{-1}
	\end{align}
	where $ \sigma_{\varrho,x_\phi} = \sigma _{\varrho,x}\cos(\phi)^2-2\sigma _{\varrho,xp}\cos(\phi)\sin(\phi)+\sigma _{\varrho,p}\sin(\phi)^2$ and analogously for $\sigma _{E,x_{\phi}}$Eq. \eqref{ee3} shows that if $\varrho$ is prepared in a state that is squeezed along $x\, (p)$ and $E$ represents a Gaussian measurement of $p\,(x)$, it is possible to retrodict the outcome values for measurements of both $x \,\mathrm{and}\, p$ with an uncertainty product that is below the HUR. This should not come as a surprise, and it is not at variance with the derivation of the HUR \eqref{hur} which explicitly concerns the prediction of the outcome of separate {\it future} measurements of either variable. It may, however, be at variance with interpretations of quantum mechanics that associate the widths of the wave function and hence of the outcome probability distributions with an actual physical extension in position and momentum of the object.
	
	Experiments on Gaussian states of large collective spin degrees of freedom have verified the reduced variances of two 
	non-commuting observables \cite{Bao} and they have confirmed the prediction in 
	\cite{Jinglei} that prior squeezing of $\hat{x}$ and posterior squeezing of $\hat{p}$ violate the HUR for retrodicted outcomes but do not warrant the squeezing of their linear combinations $\hat{x}_\phi$     
	with $\phi = \pi/4,3\pi/4$. The polar plot Fig. \ref{f1} shows the variance of the measurement outcome distributions of $x_{\phi}$. The outer black circle shows the variance for minimum uncertainty states with equal position and momentum variances according to the usual HUR, while  Eq.\eqref{ee3} is shown with butterfly shapes. For progressively stronger prior squeezing of $\hat{x}$ and posterior squeezing of $\hat{p}$, the variances along the $\phi = \pm \pi/4,\pm 3\pi/4$ directions diverge
	
	\subsection{Entanglement with an ancilla system improves retrodiction}
	
	The results shown in Fig.1 seem to exclude the possibility of initial state preparation and final measurement schemes for which the outcome of {\it any} $\hat{x}_\phi$ can be guessed with arbitrarily small uncertainty. While the prior and posterior knowledge permits only the sharp definition of two different variables there has, however, been studies showing how to ascertain the outcome of the measurement of three different spin components, and an anecdotal version of this problem goes under the name of the Mean King's problem \cite{vaidman,Mermin}. The special solution of the Mean King's problem involves the preparations of an entangled state and post-selection by projection on an entangled state of two particles. For spin $1/2$ systems, this protocol restricts the possible  outcome of the intermediate measurement of any of the cartesian components of the spin of one of the particles to definite values. 
	
	We shall show that by use of entangled Gaussian states and measurements of two oscillator modes, we can retrodict the outcome of intermediate measurements of {\it any} $\hat{x}_\phi$ for one of the modes with an uncertainty that is independent of $\phi$ and which can be made small enough that the HUR is violated by an arbitrarily large factor. 
	
	Let us consider the two-mode pure squeezed state, $\varrho=|\psi_s\rangle  \langle \psi_s|$  with
	\begin{align}\label{ss2}
	|\psi_s\rangle = e^{(\hat{a}_1^\dagger\hat{a}_2^\dagger-\hat{a}_1\hat{a}_2)s/2} |0,0\rangle = \frac{1}{\cosh (s/2)}\sum _{j=0}^{+\infty}\tanh (s/2)^{j} |j,j\rangle
	\end{align}
	where $\hat{a}^\dagger_i,\hat{a}_i,\, i=1,2 $
	are harmonic oscillator raising and lowering operators and $s$ is the squeezing parameter.
	The state $\varrho$ has a two-mode covariance matrix
	\begin{align}\label{qq}
	\bm{\sigma}_{\varrho} = \frac{1}{2}
	\begin{bmatrix}
	\cosh(s) & 0 &\sinh(s)& 0 \\
	0&\cosh(s)&0&-\sinh(s)\\
	\sinh(s)&0&\cosh(s)&0\\
	0&-\sinh(s)&0&\cosh(s)
	\end{bmatrix},
	\end{align}
	referring to the variables $(\hat{x}_1,\hat{p}_1,\hat{x}_2,\hat{p}_2)$. This state can be prepared in quantum optics by a parametric oscillator. It can also be prepared by measurement of the two observables $\hat{x}_1-\hat{x}_2$ and $\hat{p}_1+\hat{p}_2$  which hereby obtain reduced variances which ascertain the entanglement of the two systems \cite{Duan,Simon}
	
	In the same way as the state $\varrho$ can be prepared by measurement backaction, a later measurement of the observables $\hat{x}_1+\hat{x}_2$ and $\hat{p}_1-\hat{p}_2$,  correspond to an effect operator $E$ with a similar form. The random outcome values govern the displacement but not the squeezing parameter $s'$ of $E$, which is set by the measurement strength. 
	
	Knowing $\rho$ and $E$, we can obtain the probability distribution for a projective measurement of an arbitrary combination $\hat{x}_{\phi}=\hat{x}_1 \cos\phi + \hat{p}_1 \sin\phi$ for one of the systems,
	\begin{align}\label{ee1}
	\mathrm{P}_{P}(x_{\phi}) = \frac{\mathrm{Tr}_{2}\big[{}_{1}\langle x_{\phi}| \rho(t)|x_\phi\rangle_{1}\, {}_{1}\langle x_{\phi}| E(t)|x_{\phi}\rangle_{1}\big]}{\mathrm{Tr}_{2}\big[\int_{x_{\phi}'}{}_1\langle x_{\phi}'| \rho(t)|x_{\phi}'\rangle_{1}\,{}_{1}\langle x_{\phi}'| E(t)|x_{\phi}'\rangle_{1}\big]}.
	\end{align}
	The expected mean value of the measurement of $\hat{x}_{\phi}$, inferred from $\varrho$ and $E$ is \begin{align}\label{mpqs}
	\bar{x}_{\phi,\mathrm{P}} &= \big[(\bm{\sigma}_{\varrho} ^{-1}+\bm{\sigma}_{E}^{-1})^{-1}(\bm{\sigma}_{\varrho} ^{-1}\bm{\bar{r}}_{\varrho}+\bm{\sigma}_{E} ^{-1}\bm{\bar{r}}_E)\big]_{1}\nonumber\\
	&= \cos (\phi)\frac{\bar{x}^{\varrho}_{1} + \bar{x}^E_{1}}{2} +  \sin (\phi) \frac{\bar{p}^{\varrho}_{1}+\bar{p}^E_{1}}{2} \nonumber\\
	&+\frac{\sinh (s-s')}{2(1+\cosh (s-s'))}[\cos (\phi)\bar{x}_{2}'' -  \sin (\phi) \bar{p}_2''],
	\end{align} 
	where $ \bar{x}^{\varrho}_{1},\bar{p}^{\varrho}_{1}$ are the mean position and momentum in the initally prepared state $\varrho$, $ \bar{x}^E_{1},\bar{p}^E_{1}$ are the similar centroid values of the Gaussian Wigner function for $E$, determined by the final measurement, and $ \bar{x}_{2}'' = \bar{x}^E_{2}-\bar{x}^{\varrho}_{2}, \bar{p}_2'' =\bar{p}^E_{2}-\bar{p}^{\varrho}_{2}$, with similar centroid values defined for the second system.
	Eq.\eqref{mpqs} yields the most likely value and is hence the best guess for the outcome of the 
	measurement. The mean squared error of this guess is equal to the  variance of the distribution $P_P(x_\phi)$ which can be determined from,
	\begin{align}\label{s}
	\sigma _P(x_\phi) = \frac{\cosh(s) + \cosh(s')}{4(1+\cosh( s-s'))}.
	\end{align}
	Eq. \eqref{s}, shows that the variance of the retrodicted values is independent of the orientation $\phi$ in phase space and that it has no fundamental lower limit, if the density matrix and the effect matrix are equivalent to Enstein-Podolsky Rosen quantum correlated states \cite{EPR} with squeezed uncertainties in the variables $\lbrace(x_1+x_2),(p_1-p_2)\rbrace,\lbrace(x_1-x_2)$ and $(p_1+p_2)\rbrace$, respectively. 

	\section{Joint measurements of more than one quadrature  observable}\label{S:sn}
	
	In the Introduction we recalled that the textbook derivation of the HUR has nothing to do with the ability to ascertain the outcomes of sequential or simultaneous measurements of more than a single observable on the same system.
	For various sensing applications, however, it is relevant to consider the uncertainty limits for joint measurements of pairs of observables or multiple observables. In sequential measurements, the back action of the former measurement disturbs the system, and e.g., a precise position measurement leaves the system in a state with a larger momentum uncertainty. 
	
	Naturally, for an initial position squeezed state and with a final precise measurement of the momentum, it is possible to accurately predict the outcomes of an intermediate sequence of position and momentum measurements, in precisely that order. This scenario, however, seems trivial and of little metrological relevance.  
	
	\subsection{Heterodyne detection, projection on coherent state}
	
	In quantum optics it has been of interest to measure the complex field amplitude, composed of a real and an imaginary part equivalent to the position and momentum observable of a quantum particle. Such a joint measurement can be carried out by heterodyne detection which is equivalent to the projection on a coherent state $|\alpha \rangle$. The coherent states are not orthogonal but the operators $\hat{M}_{\alpha} = \frac{1}{\sqrt{\pi}}|\alpha\rangle \langle\alpha|$ form a POVM, and we can therefore identify the outcome probability distribution for any given input state $\varrho$ with the Husimi-function $Q(\alpha)=\langle\alpha|\varrho|\alpha\rangle$. The measured real and imaginary parts show fluctuations exceeding the corresponding widths of their individual projective outcome distributions (limited by the HUR), cf., the half-quantum, 3 dB added vacuum noise mentioned in Sec. II.
	
	If the system is subject to a final detection, represented by the matrix $E$, the retrodicted probability of the heterodyne measurement is
	\begin{align}\label{h2}
	P_P(x,p) \propto \langle \alpha|\varrho|\alpha\rangle \langle \alpha|E|\alpha\rangle,
	\end{align}
	the product of two Husimi functions.
	
	If both $\varrho$ and $E$ correspond to coherent states, the added noise by the heterodyne detection is thus compensated by a reduction in variance from the product of two Gaussian functions, and the retrodicted values for the joint measurement of both $x$ and $p$ then acquire the same uncertainty product as the usual HUR.
	
	
	\subsection{Joint measurements by coupling to several  meter systems} 
	
	The measurement of an observable $\hat{X}$ of a quantum system may take place by the interaction with a separate meter system with quadrature operators $(\hat{q},\hat{\pi}) $, on which a final projective measurement takes place. Thus, a model interaction of the abrupt form, $-\delta(t) \hat{X}\hat{q}$ will cause an instantaneous change of the meter momentum, $\hat{\pi}\rightarrow \hat{\pi}+\hat{X}$. If the initial quantum state of the meter has a vanishing expectation value and a very small uncertainty of the meter momentum, this will permit  a very precise and nearly projective measurement of $\hat{X}$, while an initial state with a large momentum uncertainty will yield a less precise determination of $\hat{X}$. 
	
	The narrow or broad momentum distributions conversely imply a broad or narrow position wave function $\psi(q)$ for the meter and hence a larger or smaller perturbation of the system by the system-meter interaction $\propto \hat{q}$. 
	The interplay between the measurement precision, $\propto \Delta \pi$, and the magnitude of the perturbation, $\propto \Delta q$, therefore limits the ability to predict the outcome of sequential measurements on a single quantum system. With pre- and postselection and with the entanglement with an ancillary system, we shall see, however, that this difficulty can be overcome, even for the joint measurement of a large number of non-commuting observables $\hat{x}_{\phi_j}$.   
	
\begin{figure}[ht!]
		\includegraphics[scale=.23]{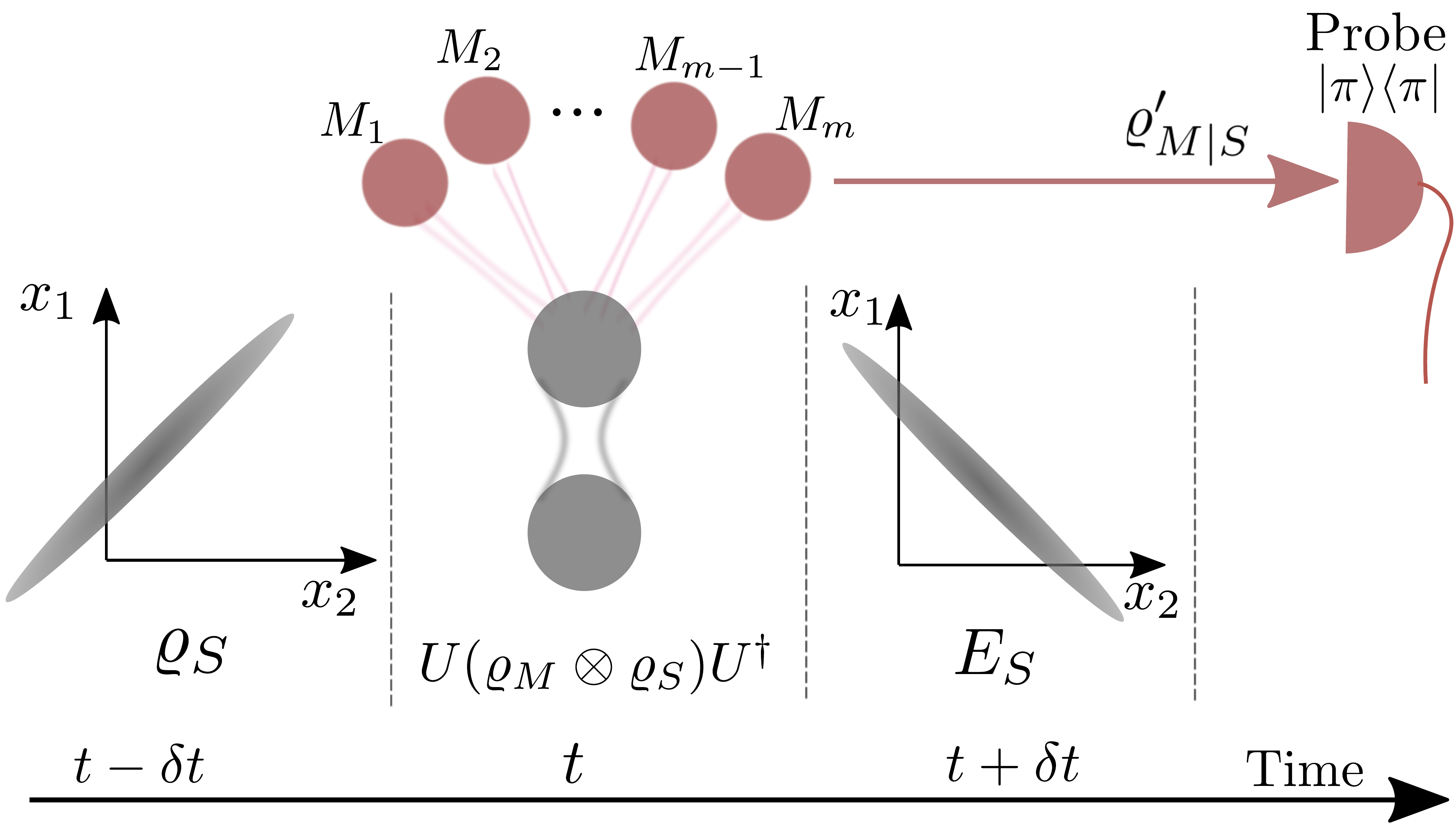}
		\caption{Scenario for the joint measurement of $m$ non-commuting quadrature operators $ \hat{x}_{\phi_i}$. $\varrho_S$ represents a two-mode squeezed state (the correlations of the $x$ quadratures are illustrated in the inset)  where one of the modes is coupled to $m$ different meter systems by the Hamiltonian \eqref{hint}. The two modes are subsequently exposed to an EPR measurement, yielding an effect operator $E_S$ of the same form as a two-mode squeeezed state with a random displacement. In this scenario it is possible to retrodict all the meter outcomes or, equivalently, predict the same values if the measurements on the meters are deferred to the end of the protocol as shown in the figure.}
		\label{f2}
	\end{figure}

	We address $m$ non-commuting quadrature operators $\hat{x}_{\phi_j}$ at equidistant orientations $\phi_j=2\pi j/m$, and with the  commutators, $[\hat{x}_{\phi_i},\hat{x}_{\phi_j}] = i\sin(\phi_j-\phi_i)$.
	A simultaneous measurement of all these observables is accomplished by simultaneously coupling the system to $m$ different meters, $ \lbrace M_1,M_2,...,M_{m-1},M_m\rbrace $, see Fig. \ref{f2}, with the interaction Hamiltonian 
	\begin{equation} \label{hint}
	H = -\delta (t) \sum_{i=1}^{m} \hat{q}_{i}\hat{x}_{\phi_i},
	\end{equation}
	where the $m$ meters have the quadrature operators  $\lbrace (\hat{q}_1,\hat{\pi}_1),(\hat{q}_2,\hat{\pi}_2),...,(\hat{q}_m,\hat{\pi}_m) \rbrace$ and they are prepared in an initial product state $\varrho _M = \varrho _{M_1}\otimes...\otimes\varrho _{M_m}$ with total covariance matrix $\bm{\sigma}_M$ and mean value vector $\hat{\bm{r}}_M$ where each meter $i$ with a Gaussian state $\varrho _{M_i}$ is specified by mean value $\bar{r}_{i}=(\bar{q}_{i},\bar{\pi}_{i})$ and covariance matrix $
	\sigma _{M_i} = \begin{bmatrix}
	\frac{z}{2} &0\\0&\frac{1}{2z}
	\end{bmatrix}$.

	The total initial state of the meters and system is considered as $\varrho_{M+S} = \varrho_M\otimes \varrho_S$ with uncorrelated covariance matrix $\bm{\sigma}_{M+S} =\begin{bmatrix}
	\bm{\sigma} _M &0\\0&\bm{\sigma}_{S}
	\end{bmatrix}$ and mean value vector $\bar{\bm{r}}_{M+S} =  \begin{bmatrix}
	\bar{\bm{r}}_{M}\\\bar{\bm{r}}_{S}
	\end{bmatrix} $. 
	In the Heisenberg picture the system operators change with the evolution operator $
	\hat{U} = e^{i\sum_{i=1}^{m} \hat{X}_{\phi _i} \hat{q}_{i}}
	$ yielding the transformation \cite{Aharonov,Botero}
	\begin{align}\
	\hat{U}^{\dagger}\hat{x}_1\hat{U}&\rightarrow \hat{x}_1 -\sum_{i=1}^{m} \sin(\phi_i) \hat{q}_i \label{b6}\\
	\hat{U}^{\dagger}\hat{p}_1\hat{U}&\rightarrow \hat{p}_1 +\sum_{i=1}^{m} \cos(\phi_i) \hat{q}_i \label{b7}
	\end{align}
	and the meter operators are governed by  
	\begin{align}
	\hat{U}^{\dagger}\bm{\hat{q}}\hat{U} &\rightarrow \bm{\hat{q}}\label{b8}\\ 
	\hat{U}^{\dagger}\bm{\hat{\pi}}\hat{U} &\rightarrow \bm{\hat{\pi}} + \bm{\hat{x}}_{\bm{\phi}} + \frac{1}{2}\bm{C}\bm{\hat{q}},\label{b9}
	\end{align}
	where the bold face symbols represent the corresponding $m$ variables in vector form, $ \bm{C}_{ij} = i[\hat{x}_{\phi_i},\hat{x}_{\phi_j}] = \sin(\phi_i-\phi_j)  $, and $\bm{\hat{x}}_{\bm{\phi}} = ( \hat{x}_{\phi_1},\hat{x}_{\phi_2},...,\hat{x}_{\phi_n} )$.

	\subsubsection{Prediction}
	The probed quantum system ($\hat{x}_1,\hat{p}_1$) is part of an EPR entangled pair, and it follows from (\ref{b9}) that the projective measurement of all the meter momentum observables yield the proper mean values and hence  constitute measurements of the system quadrature observables, $ \lbrace \hat{x}_{\phi_1},\hat{x}_{\phi_2},...,\hat{x}_{\phi_n} \rbrace$.

	Arranging the complete set of quadrature operators in a vector $ \bm{\hat{r}}_{M+S} = (\hat{q}_1,...,\hat{q}_m,\hat{\pi}_1,...,\hat{\pi}_m,\hat{x}_1,\hat{p}_1,\hat{x}_2,\hat{p}_2) $, Eqs. (\ref{b6}--\ref{b9}) specify a linear transformation, $ \bm{\hat{r}}'_{M+S} =  L\bm{\hat{r}}_{M+S} $. The mean values of these observables change by the same linear transformation while the covariance matrix $\bm{\sigma}_{M+S}$ changes by  $ \bm{\sigma}'_{M+S} = L \bm{\sigma}_{M+S} L^{\mathrm{T}}  =\begin{bmatrix}
	\bm{\sigma}' _M &\bm{\sigma}'_{MS}\\\bm{\sigma}'^{\mathrm{T}}_{MS}&\bm{\sigma}'_{S}
	\end{bmatrix} $ where the meter covariance matrix is found $\bm{\sigma}'_{M} =\begin{bmatrix}
	\bm{\sigma}' _{\bm{q}} &\bm{\sigma}'_{\bm{q\pi}}\\\bm{\sigma}'^{\mathrm{T}}_{\bm{q\pi}}&\bm{\sigma}'_{\bm{\pi}}
	\end{bmatrix}$ and the meter momentum variances conditioned only on the initial state of the systems and meters are 
	\begin{align} \label{nonpostmeter}
	(\bm{\sigma} '_{\bm{\pi}})_{ii} = \frac{1}{2z} + z\sum_{j=1}^{m}\frac{\sin(\phi_i-\phi_j)^2}{8}+\frac{\cosh (s)}{2}.
	\end{align}
	Here, we observe that the variance is the same for all meter outcomes. Squeezing of the momenta $\hat{\pi}_i$ in the initial state of the meters by the factor  $ z^{-1} $ reduces their direct contribution to the scatter of measurement outcomes, but it yields a second term proportional with $ z $ which is due to contributions from the anti-squeezed position quadrature operators $\hat{q}_i$.

	\subsubsection{Retrodicton by a final EPR measurement}
	
	If, subsequent to the interaction of one of its components with the meters, the pair of systems is subject to a final EPR measurement we may retrodict the meter outcomes with smaller uncertainty. Since the system and meter observables commute, as shown in Fig. \ref{f2}, we can  assess their retrodicted outcome distribution as the {\it prediction} of the outcomes of the deferred measurement of all these meters {\it after} the EPR measurement $E_{S}$ on the entangled system components. Such a deferred measurement argument was applied in the original derivation of Eq.\eqref{pqs} in \cite{PQS}.      
	The joint Gaussian state of the meters conditioned on the final EPR system measurements is given by mean values and a covariance matrix which are evaluated in the Appendix. The analysis presented here extends the study of infinitely sharp measurement in \cite{Aharonov,Botero} to regimes with finite squeezing resources.     
	
	The joint probability distribution of the outcomes of the momentum measurements is given by the corresponding momentum components (blocks), $\bm{\sigma}'_{\bm{\pi}|S} \equiv \bm{\sigma}_{\bm{\pi}|\varrho'_{M|S}}$ 
	and $\bar{\bm{\pi}}'_{\bm{\pi}|S} \equiv \bar{\bm{\pi}}_{\bm{\pi}|\varrho'_{M|S}}$ 
	of the matrices and vectors given in Eqs. (\ref{b5},\ref{b37}),
	\begin{align}\label{h4}
	P(\bm{\pi} | \varrho'_{M|S}) \propto
	e^{-(\bm{\pi}-\bar{\bm{\pi}}'_{\bm{\pi}|S})^{\mathrm{T}}(2\bm{\sigma}_{\bm{\pi}|S}')^{-1}(\bm{\pi}-\bar{\bm{\pi}}'_{\bm{\pi}|S})}
	\end{align}
	For $ m $ observables, the diagonal and non-diagonal elements of $\bm{\sigma}_{\bm{\pi}}$, take the form
	\begin{align}\label{s1}
	(\bm{\sigma}_{\bm{\pi}|S}')_{ii} &= \frac{1}{2z}+\frac{\cosh(s) + \cosh(s')}{4(1+\cosh( s-s'))},\\
	(\bm{\sigma}_{\bm{\pi}|S}')_{ij} &= \cos (\phi_i-\phi_j)\big[\frac{\cosh(s) + \cosh(s')}{4(1+\cosh( s-s'))}].\label{s2}
	\end{align}
	Eq. (\ref{s1}) can be written, $(\bm{\sigma}_{\bm{\pi}|S}')_{ii} = \frac{1}{2z}+\sigma_P(x_{\phi_i})$. i.e., the variance of the outcome distribution is the sum of the initial variance of the meter state and the variance of measurement outcomes for a projective measurement of {\it a single observable} on the system itself according to the PQS formalism. Eq. \eqref{s1} also shows that by choosing $s$ large and positive and $s'$ large and negative, i.e., by employing $ \varrho _S $ (and $ E_S $) that are EPR correlated with squeezed $ x_1-x_2 $ and $ p_1+p_2  $ (and $ x_1+x_2  $  and $ p_1 - p_2  $), we can make arbitrarily good predictions of all measurement outcomes. 
	
	Again, we need to assure ourselves that the predicted or retrodicted value reflects the preparation of the system, such that e.g., a change in the initial state or the final measurement outcome has the expected effect on the meter outcome. Indeed, it follows from Eq. \eqref{b37} in the Appendix, that \begin{align}\label{h5}
	(\bar{\bm{\pi}}'_{\bm{\pi}|S})_i =& \bar{\pi_i}+
	\cos (\phi_i)\frac{\bar{x}^{\varrho}_{1} + \bar{x}^E_{1}}{2} +  \sin (\phi_i) \frac{\bar{p}^{\varrho}_{1}+\bar{p}^E_{1}}{2}
	\nonumber\\
	&+\frac{\sinh (s-s')}{2(1+\cosh (s-s'))}[\cos (\phi_i)\bar{x}''_{2} -  \sin (\phi_i)\bar{p}''_{2}] \nonumber\\
	=& \,\bar{\pi}_i + \bar{x}_{\phi_i,P},
	\end{align}
	where Eq.\eqref{mpqs} was used in the last line of the derivation and  shows that the displacement of the $\pi_i$ operator is the same as the displacement retrodicted for projective measurements of each (single) system quadrature operator. 
	
	It is quite remarkable that we are able to benefit from squeezing the momentum degree of freedom $\hat{\pi}_i$ without the variance of the corresponding position  $\hat{q}_i$ contributing to the fluctuations in the measurements of the other, non-commuting observables as in Eq.\eqref{nonpostmeter}. This is a clear consequence of the post-selection. 
	
	\subsection{Even better retrodiction of joint measurements}
	
	In the previous subsection we introduced $m$ meter systems with the purpose to probe the values of $m$ non-commuting observables, and we showed how each mean outcome is in agreement with the retrodiction for individual projective measurements of each of the observables, while the variance of each measurement can be made arbitrarily small (assuming that strong squeezing and highly correlated EPR states can be prepared and measured in the laboratory). We also noted that there are correlations \eqref{s2} among the meter observables, which is suggestive that we may infer the measurement outcomes with even higher precision. 
	
	Indeed, let us assume that the optimal measurement of the observable $ \hat{x}_{\phi_k}$ is not achieved by  the outcome of measuring $\hat{\pi}_k$, but by a linear combination of such outcomes, corresponding to the operator,
	\begin{align}\label{v1}
	\tilde{\pi}_{k} = \sum_{i=1}^{m}d_i^{(k)}\hat{\pi}_i
	\end{align}
	This combination will match the correct PQS expectation value of $\hat{x}_{\phi_k}$ as long as
	\begin{align}
	\begin{cases}\sum_{i}^{m} d_i^{(k)} \cos (\phi_i - \phi_k) = 1\\
	\sum_{i}^{m} d_i^{(k)} \sin (\phi_i-\phi_k) = 0.
	\end{cases}
	\end{align}
	and the measurement outcomes will have variances given by
	\begin{align}
	\mathrm{Var}(\tilde{\pi}_{k})= \sum_{i,j=1}^{m}d_i^{(k)}	(\bm{\sigma}_{\bm{\pi}|S}')_{ij} d_j^{(k)}.
	\end{align}
	
	To minimize $\mathrm{Var}(\tilde{\pi}_{k})$ we must find the extremal eigenvector $\overrightarrow{d}$ of the matrix 
	$\bm{\sigma}_{\bm{\pi}|S}'$ with the elements  (\ref{s1},\ref{s2}). 
	
	With our assumption of equidistantly distributed  angles $\phi_i$, the solution is 
	\begin{align}
	d_i^{(k)} = \frac{2}{m} \cos (\phi_i-\phi_k)
	\end{align}
	with the result
	
	\begin{align}
	\mathrm{Var}(\tilde{\pi}_{k})= \frac{1}{mz} +\sigma_P(x_{\phi_k}).
	\end{align}
	This is an interesting and, perhaps, surprising result. In the limit of many meters $m\rightarrow\infty$, there is no need for squeezing them, and the variance of the  measurement outcomes, processed by Eq.\eqref{v1}, is the same as the variance of individual (single) projective measurements according to the PQS formalism.
	\begin{align}
	\mathrm{Var}(\tilde{\pi}_{k}) \rightarrow \sigma_P(x_{\phi_k}).
	\end{align}
	
	\section{Conclusion}\label{S:S}
	In this article we have studied the limitations set by quantum mechanics for the prediction and retrodiction of measurements of position and momentum observables and their linear combinations. We emphasize the difference between experiments that measure only one observable and the ones that aim to measure two or many observables on the same quantum system. While it is a less surprising effects that post selection enables retrodiction of the outcome of two complementary measurements, we have shown how the retrodiction capabilities are significantly enriched in measurements on parts of larger entangled systems. In such schemes it is possible to retrodict the outcome of measurements of many rather than just two different observables, and we may even retrodict the outcomes of joint measurements of many observables. 
	
	Our highly accurate retrodiction is conditioned on the outcome of the postselection measurement, but we emphasize that in our examples with Gaussian states and measurements, we do not rely on the heralding by a particular, and rare outcome of the postselection process. The final EPR measurement succeeds every time \cite{Polzik}, but from shot to shot it yields a different displacement vector that has to be invoked in the retrodiction process, cf. Eq.\eqref{mpqs}. 
	
	It is important to emphasize that the outcome of the experiments are not trivial but strictly correlated with the preparation and postselection procedure. The joint measurements by means of meter systems thus have the same retrodicted guessed values as apply for individual projective measurements of the same observables $\hat{x}_i$, cf. Eq.\eqref{mpqs}. The measurement protocols and  the calculated variances  of the outcomes of different measurement protocols hence perfectly qualify as measurements and measurement uncertainties of the said observables.  
	
	The entanglement with an ancilla system plays a special role in the Mean Kings' problem \cite{vaidman,Mermin}, and EPR states with an ancilla oscilator are also employed in back action evasion protocols for high precision sensing with a single oscillator
	\cite{woolley,caves2,moller}. We suggest to explore these ideas beyond the use of Gaussian states and also to investigate if the accomplishments obtained here by the use of EPR entanglement can be obtained with states possessing different quantum correlations such as quantum discord \cite{discord} or steering properties \cite{steering}.

	\begin{acknowledgements}
		The authors are grateful to Michael J. W. Hall for comments and suggestions for the presentation and A. T. Rezakhani for useful discussions. This  work  is  supported  by the Danish National Research Foundation through the Center of Excellence “CCQ” (Grant agreement  No.  DNRF156);  and  the  European  QuantERA grant C’MON-QSENS!, by Innovation Fund Denmark GrantNo.  9085-00002B. 
		
	\end{acknowledgements}
	
	\appendix
	\section*{Appendix : Joint Measurement by meter systems }
	\label{App1}
	The initial state of the meters and system is $\varrho_{M+S} = \varrho_M\otimes \varrho_S$ which both are considered as Gaussian state with uncorrelated covariance matrix $\bm{\sigma}_{M+S} =\begin{bmatrix}
	\bm{\sigma} _M &0\\0&\bm{\sigma}_{S}
	\end{bmatrix}$ and mean value vector $\bar{\bm{r}}_{M+S} =  \begin{bmatrix}
	\bar{\bm{r}}_{M}\\\bar{\bm{r}}_{S}
	\end{bmatrix} $. Under the interaction Hamiltonian \eqref{hint}, in the Heisenberg picture, the density matrix is transformed into $\varrho'_{M+S} = U (\varrho_M\otimes \varrho_S)U^\dagger $ with the covariance matrix $\bm{\sigma}'_{M+S} =\begin{bmatrix}
	\bm{\sigma}' _M &\bm{\sigma}'_{MS}\\\bm{\sigma}'^{\mathrm{T}}_{MS}&\bm{\sigma}'_{S}
	\end{bmatrix}$ and mean values $ \bar{\bm{r}}'_{M+S} =  \begin{bmatrix}
	\bar{\bm{r}}'_{M}\\\bar{\bm{r}}'_{S}
	\end{bmatrix} $ described in the main text.
	
	The Gaussian density matrices yield the characteristic function $\chi (\textbf{r}) = \mathrm{Tr} (\hat{\varrho} \hat{D}_{-\textbf{r}})$ and they can be expressed by and expansion on the displacement operators $\hat{D}_{\textbf{r}} $ as,
	\begin{align}\label{b3}
	\varrho'_{M+S} &= \frac{1}{(2\pi)^{m+n}}\int_{\mathbb{R}^{2(m+n)}} \mathrm{d}\textbf{r}' \chi(\textbf{r}')\hat{D}_\textbf{r}' \nonumber \\&= \frac{1}{(2\pi)^{m+n}}\int_{\mathbb{R}^{2(m+n)}} \mathrm{d}\tilde{\textbf{r}}' e^{-[\tilde{\textbf{r}}'^T \frac{\bm{\sigma}'_{M+S}}{2} \tilde{\textbf{r}}'] + i\tilde{\textbf{r}}'\bar{\textbf{r}}'_{M+S}} \hat{D}_{\Omega^T\tilde{\textbf{r}}'}
	\end{align}
	  where $i\Omega = [\hat{\bm{r}},\hat{\bm{r}}^{\mathrm{T}}]$ and $ \tilde{\textbf{r}}' = \Omega \textbf{r}'$ and the number of modes $m,n$ are corresponding to meters and system, respectively. 
	
	The conditioned state of the meters $\varrho' _{M|S}$ by which the outcome measurement of $\hat{\pi}_i$ is obtained is given by 
	$ \varrho'_{M|S} = \mathrm{Tr}_S\big[ \varrho'_{M+S} E_S \big]$. 
	By supposing the covariance matrix $\bm{\sigma}_E$ and the centroid values $\bm{\bar{r}}_E$ for the entangled system effect matrix, and by denoting the corresponding characteristic function $\chi(\bm{r}_E)$ and displacement operator $\hat{D}_{\bm{r}_E} $,  we have
	\begin{align}\label{b10}
	\varrho'_{M|S} = \mathrm{Tr}_S\big( \frac{1}{(2\pi)^{m+2n}}\int_{\mathbb{R}^{2m'}} \mathrm{d}\textbf{r}'\int_{\mathbb{R}^{2n}} \mathrm{d}\textbf{r}_E \chi(\textbf{r}')\chi(\bm{r}_E)\hat{D}_{\bm{r}_E} \hat{D}_{\textbf{r}'}  \big)
	\end{align}
	where $m' = m+n$. Using the orthogonality relation \cite{Serafini}
	\begin{align}
	\mathrm{Tr} \big[ \hat{D}_s\hat{D}_r \big] = (2\pi)^{n}\delta^{2n}(r+s)
	\end{align}
	Eq. \eqref{b10} can be rewritten,
	\begin{align}\label{b11}
	\varrho'_{M|S} = \frac{1}{(2\pi)^{m+n}}\int \mathrm{d}\tilde{\textbf{r}}'_M\mathrm{d}\tilde{\textbf{r}}'_S\Gamma (\tilde{\textbf{r}}'_M) \Gamma (\tilde{\textbf{r}}'_S) \hat{D}_{\Omega^T\tilde{\textbf{r}}'_M},
	\end{align}
	where
	\begin{align}\label{b12}
	\Gamma (\tilde{\textbf{r}}_{M}') &= e^{-[\tilde{\textbf{r}}_{M}'^{T} \frac{\bm{\sigma}_{M}'}{2} \tilde{\textbf{r}}_{M}'+\tilde{\textbf{r}}_{M}'^{T} \frac{\bm{\sigma}_{SM}'}{2} \tilde{\textbf{r}}_{S}'] + i\tilde{\textbf{r}}_{M}'\bar{\textbf{r}}_{M}'}\nonumber\\
	\Gamma (\tilde{\textbf{r}}_{S}') &= e^{-[\tilde{\textbf{r}}_{S}'^{T} \frac{\bm{\sigma}_{S}'+\bm{\sigma}_{E}}{2} \tilde{\textbf{r}}_{S}'] + i\tilde{\textbf{r}}_{S}'(\bar{\textbf{r}}_{S}'-\bar{\textbf{r}}_{E})}.
	\end{align}
	Finally, we integrate over the system degrees of freedom and obtain the conditioned state of the meters
	\begin{align}
	\varrho'_{M|S} \propto \frac{1}{(2\pi)^{m}}\int_{\mathbb{R}^{2m}} \mathrm{d}\tilde{\textbf{r}}_M e^{-[\tilde{\textbf{r}}_M^T \frac{\bm{\sigma}'_{M|S}}{2} \tilde{\textbf{r}}_M] + i\tilde{\textbf{r}}_M\bar{\textbf{r}}_{M|S}'} \hat{D}_{\Omega^T\tilde{\textbf{r}}_M}.
	\end{align}
	This expression permits extraction of the covariance matrix and the mean values for the joint state of the meters \cite{plenio} 
	\begin{align}\label{b5}
	\bm{\sigma}_{M|S}' &= \bm{\sigma}_{M}'-\bm{\sigma}_{SM}' \frac{1}{(\bm{\sigma}_{S}'+\bm{\sigma}_{E})} \bm{\sigma}_{SM}'^{\mathrm{T}}\\
	\bar{\textbf{r}}_{M|S}'&=\bar{\textbf{r}}_{M}'+\bm{\sigma}_{SM}' \frac{1}{(\bm{\sigma}_{S}'+\bm{\sigma}_{E})}(\bar{\textbf{r}}_{E}-\bar{\textbf{r}}_{S}')\label{b37}
	\end{align}
	If one considers the block covariance matrix for the meters $\bm{\sigma}'_{M|S} =\begin{bmatrix}
	\bm{\sigma}' _{\bm{q}|S} &\bm{\sigma}'_{\bm{q\pi}|S}\\\bm{\sigma}'^{\mathrm{T}}_{\bm{q\pi}|S}&\bm{\sigma}'_{\bm{\pi}|S}
	\end{bmatrix}$ and mean vector $\bar{\textbf{r}}_{M|S}' = (\bar{\bm{q}}'_{\bm{q}|S},\bar{\bm{\pi}}'_{\bm{\pi}|S}) $, the projective measurements in the basis $ |\bm{\pi} \rangle = |\pi_{1},...,\pi_{m} \rangle$ are governed by $\bm{\sigma}'_{\bm{\pi}|S} $ and $\bar{\bm{\pi}}'_{\bm{\pi}|S} $ as expressed by \eqref{h4}.

\end{document}